\documentclass[prd,preprint,superscriptaddress,amsmath,amssymb,nofootinbib]{revtex4}
\pdfoutput=1

\usepackage{graphicx,color,slashed}

\def\be{\begin{eqnarray}}
\def\ed{\end{eqnarray}}

\begin{document}


\title{\bf \Large Probing the decay $A^0\to h^0 Z^{(*)}$ in Two-Higgs-Doublet Models in the inverted hierarchy scenario \\ at the Large Hadron Collider}

\author{A.G. Akeroyd}
\email{a.g.akeroyd@soton.ac.uk}
\affiliation{School of Physics and Astronomy, University of Southampton,
Highfield, Southampton SO17 1BJ, United Kingdom}

\author{S. Alanazi}
\email{swa1a19@soton.ac.uk; swalanazi@imamu.edu.sa}
\affiliation{School of Physics and Astronomy, University of Southampton,
Highfield, Southampton SO17 1BJ, United Kingdom}
\affiliation{Physics Department, Imam Mohammad Ibn Saud Islamic University (IMISU), P.O. Box 90950, Riyadh, 11623, Saudi Arabia}

\author{Stefano Moretti}
\email{s.moretti@soton.ac.uk; stefano.moretti@physics.uu.se}
\affiliation{School of Physics and Astronomy, University of Southampton,
Highfield, Southampton SO17 1BJ, United Kingdom}
\affiliation{Department of Physics and Astronomy, Uppsala University, Box 516, SE-751 20 Uppsala, Sweden}

\date{\today}

\begin{abstract}
\noindent
Searches are being carried out at the Large Hadron Collider (LHC) for the decay of the CP-odd scalar ($A^0$) in Two-Higgs-Doublet Models (2HDMs) with Natural Flavour Conservation (NFC) in the channel
$A^0\to h^0 Z$, where $h^0$ is either the discovered 125 GeV Higgs boson or is an undiscovered CP-even scalar with a mass below 125 GeV. The latter possibility is called the "inverted hierarchy scenario" (IH) and
would provide the opportunity of simultaneous discovery of two scalars. In both searches the selection cuts are optimised for the case of an on-shell $Z$ boson. For the case of the $Z$ boson being off-shell
(denoted by $Z^*$, for which $m_{A^0} - m_{h^0}<m_Z$) no limits are set on the relevant 2HDM parameters from this process.
It is known that the decay $A^0\to h^0 Z^*$ can have a large branching ratio (BR) in 2HDMs (especially in the Type I structure).
In the context of the IH scenario and developing our previous work, we 
calculate the signal cross section $\sigma(gg\to A^0)\times {\rm BR}(A^0\to h^0Z^{(*)})\times {\rm BR}(h^0\to b\overline b,\tau\overline \tau)$ in the four types of 2HDMs with NFC.
We also suggest some selection cuts that could provide sensitivity to $A^0\to h^0 Z^*$ in the IH scenario.

\end{abstract}

\maketitle

\section{Introduction}
\noindent
The discovery of the 125 GeV boson by the ATLAS and CMS collaborations of the Large Hadron Collider (LHC)
\cite{Aad:2012tfa,Chatrchyan:2012xdj} has been followed by a period of measuring its couplings to other particles with increasing precision.
To date, all measurements of the particle's  production cross sections and branching ratios (BRs) are in very good agreement 
(within experimental error) with those of the Higgs boson of the 
Standard Model (SM) with a mass of 125 GeV. 
The boson has been observed (with a statistical significance of greater than 
$5\sigma$) to decay in five channels ($\gamma\gamma$, $ZZ$, $W^+W^-$, $\tau^+\tau^-$, and $b\overline b$) 
 and there is some evidence for the decays to $\mu^+\mu^-$ and $Z\gamma$ e.g. see \cite{ATLAS:2018kot}.
Moreover, each of the four largest
production mechanisms (gluon-gluon fusion, vector boson $(W/Z)$ 
fusion, associated production with a vector boson, and 
associated production with top quarks) has been observed for one or more of the above decay channels.
Using the full Run II data (139 fb$^{-1}$ at $\sqrt s=13$ TeV) a "signal strength"  (i.e. cross section times BR, averaged over all channels)
has been measured to be $1.02^{+0.07}_{-0.06}$ \cite{CMS:RunII} (CMS) and $1.06\pm 0.06$ \cite{ATLAS:RunII} (ATLAS), where
the SM Higgs boson is defined to have a signal strength of 1.

Despite the fact that the observed 125 GeV boson has properties that are in very good agreement with those of the  
(solitary) Higgs boson of the SM, it is possible that it is the first scalar to be discovered from a non-minimal Higgs sector in an extension of the SM i.e. the scalar potential 
could contain more than one scalar isospin doublet as well as other representations such as scalar isospin singlets/triplets. 
The non-supersymmetric Two-Higgs-Doublet Model (2HDM) \cite{Lee:1973iz,Gunion:1989we,Branco:2011iw,Wang:2022yhm} 
contains two $SU(2)_L\otimes U(1)_Y$ isospin doublets instead of just one, and has been studied for fifty years. There are different types of 2HDM, differing in how the two doublets are coupled to the fermions.
It is well known that there are shortcomings of the SM, some examples being: 
i) neutrinos are massless in the SM but neutrino oscillation experiments show that
at least two of the neutrinos are massive;
ii) there is no dark matter candidate in the SM, but cosmological observations require such a particle (or particles);
iii) there is insufficient CP violation for baryogenesis (i.e. to create a matter-antimatter imbalance in the universe).  All these issues (and others) 
can be solved (or mitigated) if there are additional scalars. 

The observation of the 125 GeV boson is consistent with many models with a non-minimal Higgs sector, because 
 a SM-like Higgs boson is predicted in part of the model's parameter space. 
As an example of this scenario, in the aforementioned
2HDM (on which we shall focus) there is an "alignment limit" in which one of the two CP-even scalars has couplings that are the same (or very close to) those of the Higgs boson of the SM. This alignment is readily obtained
if one of the two neutral CP-even scalars remains light (taken to have a mass of 125 GeV) while all of the other four scalars have masses that are much larger (called "alignment with decoupling") \cite{Bernon:2015qea}. The alignment limit can also be realised if
all scalars are of the order of the electroweak scale (called "alignment without decoupling") and it is on this scenario that we will focus  \cite{Bernon:2015qea,Bernon:2015wef}.

Current and future runs of the LHC experiment (and proposed experiments such as an $e^+e^-$ collider) have an opportunity to determine 
whether or not the 125 GeV boson is the first scalar to be discovered from a non-minimal Higgs sector.
More precise measurements of the production cross sections and BRs might start to show deviations from the values for the SM Higgs boson of 125 GeV.
In addition,
non-minimal Higgs sectors contain additional electrically neutral scalars and/or charged scalars, and discovery of a second Higgs boson would be unequivocal evidence  of 
such a structure. In 2HDMs there are five Higgs bosons in total: two CP-even scalars $h^0$ and $H^0$ (with $m_{h^0} < m_{H^0}$), a pair of charged scalars $H^+$ and $H^-$ and
a neutral pseudoscalar Higgs boson $A^0$, which is CP-odd. All of these particles are being searched for at the LHC.

In the context of a 2HDM, the discovered 125 GeV boson (which has been shown to be CP-even) has two interpretations. It is either the lighter CP-even $h^0$ 
(this scenario is called "normal hierarchy", NH) or the heavier CP-even $H^0$ (a scenario called "inverted hierarchy", IH).
The CP-odd $A^0$, although being electrically neutral, cannot be interpreted as the 125 GeV boson due to its different phenomenology to both $h^0$, $H^0$ and the SM Higgs boson (e.g. it does not have tree-level couplings to $W^\pm$ and $Z$).
In this paper we focus on the prospects of discovering an $A^0$ from a 2HDM at the LHC via its decay $A^0\to h^0Z^{(*)}$. Although other ways of discovering $A^0$ are possible (e.g. via its decays to fermions), the
decay $A^0\to h^0Z^{(*)}$ is particularly appealing in the IH scenario as it would allow discovery of two Higgs bosons (i.e. $A^0$ and $h^0$, where $m_{h^0}<125$ GeV).
In the context of the NH scenario, the decay $A^0\to h^0Z^{(*)}$ involves the discovered 125 GeV boson $h^0$, while observation of the related decay $A^0\to H^0 Z$ would lead to the discovery of two Higgs bosons.

Searches have been carried out for $A^0\to h^0Z$ and $A^0\to H^0Z$ (assuming an on-shell $Z$) at the LHC
\cite{ATLAS:2015kpj,CMS:2015uzk,ATLAS:2017xel,CMS:2019qcx,CMS:2019kca,CMS:2019ogx,ATLAS:2020gxx,ATLAS:2022enb,ATLAS:2023zkt}.
In a previous work \cite{Akeroyd:2023kek} we quantified the magnitude of $\sigma(gg\to A^0)\times {\rm BR}(A^0\to h^0Z^{(*)})\times {\rm BR}(h^0\to b\overline b)$ in the
2HDM (Type I and Type II), showing that the cross section can be much larger in IH than in NH\footnote{In Ref.~\cite{Akeroyd:2024dzq} the interference of the signal
$gg\to A^0\to h^0Z^{(*)}\to h^0 l^+l^-$ and SM processes that lead to the same final state was studied.}.
However, we erroneously stated in  \cite{Akeroyd:2023kek}  that there had been no search at the LHC for $A^0\to h^0 Z$ that is sensitive to $m_{h^0}<125$ GeV.
Since the writing of  \cite{Akeroyd:2023kek}  we became aware of the search in \cite{CMS:2019ogx} which does probe the region of $m_{h^0}<125$ GeV. Moreover, we became aware of the work 
\cite{Bernon:2015wef} which also studied the magnitude of $\sigma(gg\to A^0)\times {\rm BR}(A^0\to h^0Z^{(*)})\times {\rm BR}(h^0\to b\overline b)$ in 2HDMs in the IH scenario. In this work 
we will develop and improve our previous work in \cite{Akeroyd:2023kek} in the following three ways: i) we describe the search in 
\cite{CMS:2019ogx}, studying its implications for our previous and present work, and emphasising that there are no limits in \cite{CMS:2019ogx} on the decay $A^0\to h^0Z^*$ (i.e. $Z$ is off-shell) although its BR can be large in 2HDMs; ii) clarify the overlap and
differences between our work \cite{Akeroyd:2023kek} and that of \cite{Bernon:2015wef}, extending the work of \cite{Bernon:2015wef} to include two other 2HDMs, and updating their results (from 2015) by using the most up-to-date experimental constraints;
 iii) to suggest selection cuts that could help gain sensitivity to the decay $A^0\to h^0Z^*$ and
improve the detection prospects for this decay channel.
 
The structure of this work is as follows. In section II the various forms of 2HDM are introduced as well as theoretical and experimental constraints on its parameters.
In section III a brief introduction to the decay $A^0\to h^0 Z^{(*)}$ is given and in section IV the searches for $A^0\to h^0 Z$ at the LHC are described.
Our numerical results for the cross section for $gg\to A^0\to h^0 Z^{(*)}$ events in the IH scenario are presented in section V, and conclusions are given in
section VI.

\section{The Two Higgs Doublet Model (2HDM)}
\noindent
The SM contains one complex scalar isospin doublet $(I=1/2)$ with hypercharge $Y=1$. The minimum of the scalar potential is obtained at a non-zero value of the real part of the neutral scalar field, which is referred to 
as a vacuum expectation value ($v$).
The presence of $v$ causes the spontaneous breaking of the $SU(2)_L\otimes U(1)_Y$ local gauge symmetry to a $U(1)_Q$ local gauge symmetry ("electroweak symmetry breaking" (EWSB)),
thus providing mass to the $W^\pm, Z$ (via the kinetic energy term of the scalar fields)
and charged fermions (via the Yukawa couplings). This way of generating mass is called the "Higgs mechanism", and gives rise to a CP-even physical scalar particle (a "Higgs boson", $h^0$). In the context of the SM the Higgs boson $h^0$ has now been 
discovered with a mass of around 125 GeV.
 The Higgs mechanism can also be implemented in the 2HDM  \cite{Lee:1973iz,Gunion:1989we,Branco:2011iw,Wang:2022yhm}
 in which each of the complex scalar doublets has a vacuum expectation value ($v_1$ and $v_2$).
The 2HDM has been well-studied as a minimal extension of the SM and such a structure is also necessary in supersymmetric (SUSY) extensions of the SM.
 As mentioned in the introduction, in 2HDMs there are five physical Higgs bosons:  two CP-even scalars $h^0$ and $H^0$ (with $m_{h^0} < m_{H^0}$), a pair of charged scalars $H^+$ and $H^-$ and
a neutral pseudoscalar Higgs boson $A^0$, which is CP-odd. 
 In the context of a 2HDM the discovery of the 125 GeV boson at the LHC is interpreted as being either
  $h^0$ (NH scenario) or $H^0$ (IH scenario), with couplings very close to those of the SM Higgs boson.

The scalar structure of 2HDMs (being more complicated than that in the SM) is restricted in various ways.
In 2HDMs the neutral scalars are sources of tree-level "Flavour Changing Neutral Currents" (FCNCs) which must be strongly suppressed in order to comply with experiment. 
 Such FCNCs lead to interactions that change quark flavour (such as a vertex $h^0b\overline s$).
A simple suppression mechanism of FCNCs in 2HDMs (referred to as the "Paschos-Glashow-Weinberg theorem" or Natural Flavour Conservation (NFC) \cite{Glashow:1976nt}) is to require that the Lagrangian respects certain discrete symmetries ($Z_2$ symmetries).
These symmetries ensure that a given flavour of charged fermion receives its mass from just one vacuum expectation value, which eliminates FCNC processes at the tree-level.

The  most general scalar potential of a 2HDM that is invariant under the $SU(2)_L\otimes U(1)_Y$ local gauge symmetry and which only softly breaks (via the $m^2_{12}$ terms) an appropriate $Z_2$ symmetry (imposed to avoid FCNCs) is as 
follows \cite{Gunion:1989we,Branco:2011iw}:
 \begin{eqnarray}
        V(\Phi _{1}\Phi _{2})  =  m_{11}^{2}\Phi _{1}^{\dagger }\Phi _{1}+m_{22}^{2}\Phi _{2}^{\dagger }\Phi _{2}-  m_{12}^{2}(\Phi _{1}^{\dagger }\Phi _{2}+\Phi _{2}^{\dagger }\Phi _{1})+  \frac{ \lambda _{1}}{2}(\Phi _{1}^{\dagger }\Phi _{1})^{2}+\\ \nonumber
         \frac{ \lambda _{2}}{2}(\Phi _{2}^{\dagger }\Phi _{2})^{2}+  \lambda _{3}\Phi _{1}^{\dagger }\Phi _{1}\Phi _{2}^{\dagger }\Phi _{2}+
          \lambda _{4}\Phi _{1}^{\dagger }\Phi _{2}\Phi _{2}^{\dagger }\Phi _{1}+\frac{ \lambda _{5}}{2}[(\Phi _{1}^{\dagger }\Phi _{2})^{2}+(\Phi _{2}^{\dagger }\Phi _{1})^{2}]\,,
 \end{eqnarray}
with $\Phi _{i}=\binom{\Phi _{i}^{\dotplus }}{\frac{(\upsilon _{i}+\rho _{i}+i\eta _{i})}{\sqrt{2}}}$,  $\rho_i$ and $\eta_i$ are real scalar fields $\:{\rm and} \; i=1,2$. \\
Some of the parameters in the scalar potential can be complex and thus be sources of CP violation. It is common in phenomenological studies of the 2HDM to consider a simplified scenario by taking all parameters to be real, and this is the 
case we will focus on.
The scalar potential then has 8 real independent parameters: $m^2_{11}$, $m^2_{22}$, $m^2_{12}$, $\lambda _{1}$, $\lambda _{2}$, $\lambda _{3}$, $\lambda _{4}$, and $\lambda _{5}$.
These parameters determine the masses of the Higgs bosons and enter the expressions for the couplings to fermions and gauge bosons. However, it is convenient to use different independent parameters which are more directly related to physical observables.
A common choice is: $m_{h^0}$, $m_{H^0}$, $m_{H^\pm}$, $m_{A^0}$, $\upsilon_1$, $\upsilon_2$, $m^2_{12}$ and $\sin(\beta-\alpha)$. The first four parameters are the masses of the physical Higgs bosons.
The vacuum expectation values $\upsilon_1$ and $\upsilon_2$ are the values of the neutral CP-even fields in $\Phi_1$ and $\Phi_2$ respectively at the minimum of the scalar potential:
\begin{equation}
    \left<\Phi _{1} \right> =\frac{1}{\sqrt{2}}\binom{0}{\upsilon _{1}}\, ,\, \, \, \, \left<\Phi _{2} \right> =\frac{1}{\sqrt{2}}\binom{0}{\upsilon _{2}}\,.
\end{equation}
The parameter $\beta$ is defined via $\tan\beta=\upsilon_2/\upsilon_1$ with $0\le \beta \le \pi/2$, while the angle $\alpha$ ($-\pi/2 \le \alpha \le \pi/2 $) determines the composition of the CP-even mass eigenstates $h^0$ and $H^0$ in terms of the original 
neutral CP-even fields that are present in the isospin doublets $\Phi_1$ and $\Phi_2$. Of these 8 independent parameters in the scalar potential, 2 have now been measured. After EWSB in a 2HDM, the mass of the $W^\pm$ boson is given by
$m_W=gv/2$, with $ \upsilon=\sqrt{\upsilon^{2}_{1}+\upsilon^{2}_{2}}\simeq 246$ GeV. Consequently, only one of $\upsilon_1$ and $\upsilon_2$ is independent, and so $\tan\beta=\upsilon_2/\upsilon_1$ is taken as
an independent parameter. As mentioned earlier, in a 2HDM the observed 125 GeV boson is interpreted as being $h^0$ or $H^0$ and thus either $m_{h^0}=125$ GeV (NH) or $m_{H^0}=125$ GeV (IH).
The remaining 6 independent parameters in the 2HDM scalar potential are: $m_{H^\pm}$, $m_{A^0}$, $m^2_{12}$, $\tan\beta$, $\sin(\beta-\alpha)$ and one of $[m_{h^0}, m_{H^0}]$. In the NH scenario $m_{H^0}>125$ GeV
and in the IH scenario $m_{h^0}<125$ GeV. In this work we will focus on the IH scenario.

As mentioned above, the  masses of the pseudoscalar $A^0$ and the charged scalars $H^{\pm}$ are independent parameters. In terms of the original parameters in 
scalar potential these masses are given at lowest order by:
   \begin{equation}
       \begin{aligned}
         m_{A^0}^{2} & = \left [\frac{ m_{12}^{2}}{\upsilon _{1}\upsilon _{2}}  -2\lambda_{5}\right ](\upsilon _{1}^{2}+\upsilon _{2}^{2})\,,\\ m_{H^{\pm }}^{2} & = \left [\frac{ m_{12}^{2}}{\upsilon _{1}\upsilon _{2}} -\lambda_{4} -\lambda_{5}\right ](\upsilon _{1}^{2}+\upsilon _{2}^{2}) = \left [ m_{A}^{2}  +\upsilon (\lambda_{5}-\lambda_{4})\right ]  \,.
       \end{aligned}
   \end{equation}
 It is evident from these equations that the mass difference between $m_{A^0}$ and $m_{H^\pm}$ depends on $\lambda_5-\lambda_4$. In our numerical analysis we
  shall take $m_{A^0}=m_{H^\pm}$ (which corresponds to $\lambda_5=\lambda_4$) in order to satisfy more easily the constraints from electroweak precision observables ("oblique parameters", to be discussed later
  in this section).
     
 We focus on 2HDMs with NFC, of which there are four distinct types. These are called: Type I, Type II, Lepton Specific and Flipped \cite{Barger}.
 Each of these 2HDMs has the same scalar potential, but differs in how the two doublets are coupled to the charged fermions.
 Each of the four models has a distinct phenomenology, which has been studied in great detail (for a review see \cite{Branco:2011iw}). In this work we shall be studying the decay $A^0\to h^0 Z^{(*)}$ and thus we give the Lagrangian for the Yukawa
 couplings of $A^0$ in the four models \cite{Branco:2011iw}:
\begin{equation}
{\cal L}^{yuk}_{A^0} =\frac{i}{v}\left(y^d_{A^0} m_d A^0 \overline d \gamma_5 d + y^u_{A^0} m_u A^0 \overline u \gamma_5 u+  y ^\ell_{A^0} m_\ell  A^0 \overline\ell \gamma_5 \ell \right)\,.
\label{yukawa}
\end{equation}
In eq.~(\ref{yukawa}) it is understood that $d$ refers to the down-type quarks ($d$, $s$, $b$), $u$ refers to the up-type quarks ($u$, $c$, $t$) and $\ell$ refers to the
charged leptons ($e$, $\mu$, $\tau$) i.e. there are three terms of the form $y^d_{A^0} m_d \overline d \gamma_5 d$.
 In Table \ref{2HDMAcoup} the couplings $y^d_{A^0}$, $y^u_{A^0}$, and $y^\ell_{A^0}$  of $A^0$ to the charged fermions in each of these four models
are displayed.
\begin{table}[h]
\begin{center}
\begin{tabular}{|c||c|c|c|}
\hline
& $y^d_{A^0}$ &  $y^u_{A^0}$ &  $y^\ell_{A^0}$ \\ \hline
Type I
&  $-\cot\beta$ & $\cot\beta$ & $-\cot\beta$ \\
Type II
& $\tan\beta$ & $\cot\beta$ & $\tan\beta$ \\
Lepton Specific
& $-\cot\beta$ & $\cot\beta$ & $\tan\beta$ \\
Flipped
& $\tan\beta$ & $\cot\beta$ & $-\cot\beta$ \\
\hline
\end{tabular}
\end{center}
\caption{The couplings $y^d_{A^0}$, $y^u_{A^0}$, and $y^\ell_{A^0}$  in the Yukawa interactions of $A^0$ in the four versions of the 2HDM with NFC.}
\label{2HDMAcoup}
\end{table}

The allowed parameter space in a 2HDM is that which respects all theoretical and experimental constraints. These constraints are listed below  \cite{Branco:2011iw}:
\begin{enumerate}
    \item {\bf Theoretical constraints}
    \begin{enumerate}
        \item[{(i)}] {\underline{Vacuum stability of the 2HDM scalar potential}}\\
         The values of the quartic couplings $\lambda_i$ are constrained by the requirement
        that the scalar potential has the following properties:\\
         a) it breaks the electroweak symmetry $SU(2)_L\otimes U(1)_Y$
        to $U(1)_Q$;\\
         b) the scalar potential is bounded from below;\\
         c) the scalar potential  stays positive for arbitrarily large values of the scalar fields.  The above requirements lead to the following constraints:\\
         $\lambda _{1}> 0, \;\;\lambda _{2}> 0, \;\; \lambda _{3}+\lambda _{4}-\left|\lambda _{5} \right| + \sqrt{\lambda _{1} \lambda _{2}} \ge 0, \;\;\lambda _{3}+ \sqrt{\lambda _{1}\lambda _{2}}\ge 0$.\\
         It can be seen that $\lambda_1$ and $\lambda_2$ are positive definite, while $\lambda_3,  \lambda_4$ and $\lambda_5$ can have either sign. 
        \item[{(ii)}] {\underline{Perturbativity}}\\
         For calculational purposes it is required that the quartic couplings $\lambda _{i}$ do not take numerical values for which the perturbative expansion  ceases to converge.
        The condition $\left| \lambda  _{i}\right|\leq 8\pi$ ensures that the couplings $\lambda _{i}$ remain perturbative up to the Planck scale.\\
        \item[{(iii)}]  {\underline{Unitarity}}\\
        The  $2\to 2$ scattering processes ($s_1s_2\to s_3s_4$) that only involve scalars (including Goldstone bosons) are mediated by scalar quartic couplings, which depend on the parameters of the scalar potential. 
        Tree-level unitarity constraints require that the eigenvalues of the scattering matrix of the amplitudes of $s_1s_2\to s_3s_4$ be less than the unitarity limit of $8\pi$, which leads to further constraints on $\lambda_i$.
    \end{enumerate}
    \item {\bf Experimental constraints}
     \begin{enumerate}
     \item[{(i)}] 
    {\underline{Direct searches for Higgs bosons}}\\
    The discovery of the 125 GeV boson at the LHC and the non-observation of other Higgs bosons at LEP, Tevatron and LHC exclude regions of the parameter space of any 2HDM.    
    These constraints are respected in our numerical results by using the publicly available code HiggsTools \cite{Bahl:2022igd}. Any point in the parameter space of a 2HDM 
    parameter that violates experimental limits/measurements concerning Higgs bosons is rejected. 
 \item[{(ii)}] {\underline{Oblique parameters}}\\
 The Higgs bosons in a 2HDM contribute to the self-energies of the $W^\pm$ and $Z$ bosons.  The so-called "oblique parameters" $S$, $T$ and $U$ \cite{Peskin:1990zt} parametrise the
 deviation from the SM prediction of $S=T=U=0$. The current best-fit values (not including the recent CDF measurement of $m_W$ \cite{CDF:2022hxs}) are \cite{ParticleDataGroup:2022pth}:
 \begin{equation}
 S=-0.01\pm 0.10,\;\; T=0.03\pm 0.12,\;\; U=0.02\pm 0.11 \,.
 \end{equation}
 
If $U=0$ is taken (which is approximately true in any 2HDM) then the experimental allowed ranges for $S$ and $T$ are narrowed to \cite{ParticleDataGroup:2022pth}:
 \begin{equation}
 S=0.00\pm 0.07,\;\; T=0.05\pm 0.06\,.
 \label{ST}
 \end{equation}
In our numerical results the theoretical constraints in 1(i), 1(ii), 1(iii) and the experimental constraints 2(ii) 
(using the allowed ranges for $S$ and $T$ in eq.(\ref{ST})) are respected by using 2HDMC \cite{Eriksson:2009ws}. 
If the recent measurement of $m_W$ by the CDF collaboration \cite{CDF:2022hxs} is included in the world average for $m_W$
 then the central values of the $S$ and $T$ parameters in eq.(\ref{ST}) alter significantly, but can be accommodated in a 2HDM
 in the case of sizeable mass splittings among the Higgs bosons - for recent studies in both NH and IH see
\cite{Abouabid:2022lpg,Lee:2022gyf}.

\item[{(iii)}] {\underline{Flavour constraints}}\\
The parameter space of a 2HDM is also constrained by flavour observables, in particular from the decays of $b$ quarks (which are confined inside $B$ mesons).
The main source of these constraints is the fact that the charged Higgs boson $H^\pm$ contributes to processes that are mediated by a $W^\pm$, which in turn leads to constraints on the
2HDM parameters $m_{H^\pm}$ and $\tan\beta$. A flavour observable that is especially constraining is the rare decay $b\to s\gamma$ \cite{Ciuchini1,Ciuchini2, Borzumati, Gambino, Misiak,Misiak2}, although $H^\pm$ contributes to numerous
other processes (e.g. $B\overline B$ mixing). Many studies of flavour constraints on the parameter space of 2HDMs have been carried out  e.g.
\cite{Arbey:2017gmh,Atkinson:2022pcn,Cheung:2022ndq}.
In our numerical analysis we respect all flavour constraints by use of the publicly available code SuperIso \cite{Mahmoudi:2008tp}.
In the 2HDM (Type I), in which the couplings of $H^\pm$ to the fermions are proportional to $\cot\beta$, the constraint on $m_{H^\pm}$ weakens with increasing $\tan\beta$.
The lowest value of  $\tan\beta$ that we consider is $\tan\beta=3$, for which $m_{H^\pm}=140$ GeV is allowed (as can be seen in \cite{Arbey:2017gmh}). In contrast, the 2HDM (Type II)
the decay  $b\to s\gamma$ leads to the bound $m_{H^\pm}>580$ GeV, irrespective of the value of $\tan\beta$.

 \end{enumerate}
  \end{enumerate}

\section{The decay $A^0\to h^0 Z^{(*)}$}
\noindent
In this section we briefly introduce the phenomenology of $A^0$ at the LHC with particular attention given to the decay $A^0\to h^0 Z^{(*)}$, which is the channel of interest in this work.
For a more comprehensive review we refer the reader to our previous work \cite{Akeroyd:2023kek}.

We now describe the main decay channels of $A^0$. Expressions for the partial width for $A^0\to f\overline f$ to a fermion ($f$) and an anti-fermion ($\overline f$) can be found in 
e.g. \cite{Djouadi:1995gv,Djouadi:2005gj,Branco:2011iw,Choi:2021nql}. These partial widths depend on the Yukawa couplings,  which differ in each the four 2HDMs as shown in Table I.
The partial width for the decay to two gluons ($A^0\to gg$) at leading order proceeds via triangle loops of fermions. The largest contributions come from: i) the triangle diagram with $t$-quarks, which is proportional to $(y^t_{A^0})^2$, and ii) the 
triangle diagram with 
$b$-quarks, which is proportional to $(y^b_{A^0})^2$.
The explicit formula for $\Gamma(A^0\to gg)$ can be found in \cite{Djouadi:1995gv,Djouadi:2005gj,Spira:1995rr}.
The decay $A^0\to \gamma\gamma$, which is mediated by triangle loops of $f$, $W^\pm$ and $H^\pm$, has a 
much smaller partial width than $\Gamma(A^0\to gg)$ because the former has a factor of $\alpha^2$ while the latter has a factor of $\alpha^2_s$ and a colour factor enhancement.
Notably, due to the CP-odd nature of $A^0$, the decays $A^0\to W^+W^-$ and $A^0\to ZZ$ are absent at tree-level in the (CP-conserving) 2HDM. These decays are generated at higher orders
but have much smaller BRs \cite{Bernreuther:2010uw,Arhrib:2018pdi} than some of the tree-level decays and will be neglected in our study.

We now discuss the decays of $A^0$ to another Higgs boson and to a vector boson. These
interactions originate  from the kinetic term in the Lagrangian and hence have no dependence on the Yukawa couplings. We will not consider the decay $A^0\to H^\pm W^\mp$ as we will take
$m_{A^0}=m_{H^\pm}$ in our study.
The partial width for $A^0\to h^0 Z$ (i.e. a two-body decay with on-shell $Z$) is given by:
\begin{equation}
\Gamma(A^0\to h^0 Z)=\frac{m^4_Z\cos^2(\beta-\alpha)}{16\pi^2 m_{A^0}v^2}\lambda\left(\frac{m^2_{A^0}}{m^2_{Z}}, \frac{m^2_{h^0}}{m^2_{Z}}  
\right)   \lambda^{1/2}\left(\frac{m^2_Z}{m^2_{A^0}}, \frac{m^2_{h^0}}{m^2_{A^0}}  \right)  \,.
\label{width_AhZ}
\end{equation}
The partial width $\Gamma(A^0 \to H^0Z)$ has the same form as eq.~(\ref{width_AhZ}), but with $m_{h^0}$ replaced by $m_{H^0}$ and
$\cos^2(\beta-\alpha)$ replaced by $\sin^2(\beta-\alpha)$.
We will be particularly interested in the decay $A^0\to h^0 Z^*$ with (off-shell) $Z^*\to f\overline f$.
The partial width $\Gamma(A^0\to h^0 Z^*)$ 
is also proportional to $\cos^2(\beta-\alpha)$, as in eq.~(\ref{width_AhZ}), due to the $A^0Zh^0$ coupling. However, in difference to eq.~(\ref{width_AhZ}) the phase space is now 3-body ($h^0f\overline f$) instead
of 2-body ($h^0Z$) and involves an integration over the momenta of $f\overline f$. The  explicit expression for $\Gamma(A^0\to h^0 Z^*)$  is given in  \cite{Djouadi:1995gv,Moretti:1994ds,Aiko:2022gmz}.

We now discuss previous studies of  $A^0\to h^0 Z^{(*)}$ in models with two Higgs doublets.
In the Minimal Supersymmetric Standard Model (MSSM), which has a 2HDM scalar potential but with fewer free parameters, BR$(A^0\to h^0 Z^{(*)})$  is always relatively small. This is because 
the value of $\sin^2(\beta-\alpha)$ rapidly approaches 1 as $m_{A^0}$ increases above 100 GeV and so $\cos^2(\beta-\alpha)\to 0$. In contrast, in a non-SUSY 2HDM
the parameter $\sin^2(\beta-\alpha)$ is independent and could differ substantially from 1 for $m_{A^0}>100$ GeV. Early studies of BR($A^0\to h^0 Z$) in the MSSM and its detection prospects at the LHC can be found in \cite{Gunion:1991cw, Baer:1992uu,Abdullin:1996as}.
The first calculation of $\Gamma(A^0\to h^0 Z^*)$ was carried out in \cite{Djouadi:1995gv,Moretti:1994ds}. 
The BRs of $A^0$ in the MSSM are summarised in \cite{Djouadi:2005gj}. For low $\tan\beta$ (e.g. $ \tan\beta=3$), BR$(A^0\to h^0 Z^{(*)})$ can be of the order of 10\% or more in the region $200\,{\rm GeV} < m_{A^0} < 300$ GeV when the
two-body decay is open and before $A^0\to t\overline t$ becomes dominant for heavier $m_{A^0}$.

In the context of non-supersymmetric 2HDMs with NFC (on which we focus) an early study of the on-shell decay $A^0\to h^0 Z$ (Type I and Type II only) was performed in \cite{Kominis:1994fa}.
 It was shown that this decay channel for $A^0$ can have the largest BR. The three-body decay $A^0\to h^0Z^*$ in non-supersymmetric 2HDMs (all four types) with NFC was first studied in the context of LEP2 in \cite{Akeroyd:1998dt}. It was pointed out
that BR($A^0\to h^0Z^*)$ can be dominant (and even approach $100\%$) in Type I as $\tan\beta$ increases because the competing partial widths $\Gamma(A^0\to f\overline f)$  $\propto \cot^2\beta$. This contrasts with the case in the MSSM  
(with necessarily Type II structure) for which BR($A^0\to h^0Z^*)$ is always small.
In \cite{Akeroyd:1998dt}, BR($A^0\to h^0Z^*)$ was studied as a function of $\tan\beta$ in the 2HDM (Type I) for $m_{A^0}=80$ GeV, 100 GeV and 120 GeV, with $m_{h^0}=40$ GeV and $\cos^2(\beta-\alpha)=1$. These parameters
correspond to the IH scenario but at that time $m_{H^0}$ was not known to be 125 GeV. More recent works on the magnitude of the BRs of $A^0$ in the four versions of the 2HDM with NFC 
are \cite{Aoki:2009ha, Aiko:2022gmz}.

The decay $A^0\to h^0Z^*$ is relatively more important in IH than in NH due to: i)  $\cos(\beta-\alpha)$ taking very different values (described below) in the two scenarios, and ii) the larger mass splitting between $m_{A^0}$ and $m_{h^0}$ in IH.
In the case of NH one has (by definition) $m_{h^0}=125$ GeV and so necessarily $m_{H^0}>125$ GeV. 
The discovered 125 GeV boson (i.e. $h^0$ in the case of NH) has been measured by the LHC experiments to have SM-like Higgs boson couplings within experimental error, and in the context of a 2HDM with NH
 the parameter $|\cos(\beta-\alpha)|$ is thus constrained to be (approximately) less than 0.1.
Since the coupling $A^0 h^0 Z$ is proportional to $\cos(\beta-\alpha)$, in NH the
 decay channel $A^0\to h^0 Z$ has a suppression factor of $|\cos(\beta-\alpha)|^2\approx 0.01$. Studies of BR$(A^0\to h^0 Z$) in this scenario are in \cite{Aiko:2020ksl,Cheung:2022ndq,Aiko:2022gmz,Accomando:2020vbo}.
 
 In the case of IH one has $m_{H^0}=125$ GeV and so necessarily $m_{h^0}< 125$ GeV. The above constraint on $\cos(\beta-\alpha)$ now applies to 
 $\sin(\beta-\alpha)$, and so $0.9 < |\cos(\beta-\alpha)|< 1$. Consequently, the decay $A^0\to h^0 Z$ has very little suppression from the coupling $A^0 h^0 Z$, in contrast to the case
 of NH. Moreover, since $m_{h^0}<125$ GeV the decay $A^0\to h^0Z$ can proceed via an on-shell $Z$ for lighter values of $m_{A^0}$ than in the case of NH
 i.e.  if $m_{h^0}=125$ GeV then $m_{A^0} > 216$ GeV is required for on-shell $A^0\to h^0Z$ but for $m_{h^0}=90$ GeV (say) then the on-shell decay $A^0\to h^0Z$ is open for $m_{A^0}> 180$ GeV.
 Moreover, off-shell decays $A^0\to h^0Z^*$ can also be dominant in the 2HDM (Type I) over a large region of parameter space of the model. 
 Previous studies of BR($A^0\to h^0Z^{(*)}$) in non-SUSY 2HDMs in IH are the aforementioned early study in \cite{Akeroyd:1998dt} (for 80 GeV $< m_{A^0}< 120$ GeV) and more recently in  
 \cite{Abouabid:2022lpg} and \cite{Moretti:2022fot}. 
 
 At the LHC the main production process (especially at low $\tan\beta$) for $A^0$ is (gluon-gluon fusion) $gg\to A^0$ 
 \cite{Djouadi:2005gj,Spira:1995rr,Bagnaschi:2022dqz} which proceeds via a top-quark loop and a bottom-quark loop, and thus involves the Yukawa couplings for the vertices $A^0 t\overline t$ and $A^0 b\overline b$. 
 In our previous work in  \cite{Akeroyd:2023kek} we studied (as a function of the 2HDM parameters) the event number 
 \begin{equation}
\sigma(gg\to A^0\to b\overline b Z^{(*)})=\sigma(gg\to A^0)\times {\rm BR}(A^0\to h^0Z^{(*)})\times {\rm BR}(h^0\to b\overline b)\,.
\label{event_number}
\end{equation}
 The focus was on the 2HDM (Type I), which gives rise to the largest cross sections. At the time of writing of  \cite{Akeroyd:2023kek}
 we were unaware of the paper \cite{Bernon:2015wef} in which $\sigma(gg\to A^0\to b\overline b Z^{(*)})$ in the 2HDM (Types I and II) in the IH scenario was also studied.
 In section V we will clarify the overlap and differences in the studies in \cite{Bernon:2015wef} and \cite{Akeroyd:2023kek}, and in this work we will develop and extend those results.

\section{Searches for $A^0\to h^0 Z$ at the LHC}
\noindent
The decay $A^0\to h^0 Z$ (and $A^0\to H^0 Z$) in which the $Z$ boson is taken to be on-shell and $h^0/H^0\to b\overline b$, has been searched for
at the LHC by the CMS  collaboration. In Ref.~\cite{CMS:2019qcx} the search strategy focusses on the discovered 125 GeV Higgs boson, which is taken to be $h^0$ in NH and $H^0$ in IH.
The search is optimised for the case of an on-shell $Z$ boson by employing a cut on the invariant mass of the leptons (chosen as 70 GeV$ < m_{l^+l^-} < 110$ GeV) that arise from the decay $Z\to l^+l^-$.
To improve the signal to background ratio, a cut on the invariant mass of the $b\overline b$ pair arising from $h^0/H^0\to b\overline b$ was employed in order to take advantage of the fact that 
$m_{h^0}=125$ GeV in NH and $m_{H^0}=125$ GeV in IH.
Limits on the 2HDM parameters $[m_{A^0}, \tan\beta, \cos(\beta-\alpha)]$ in the case of NH  were presented for $m_{A^0} \ge 225$ GeV i.e. the $Z$ boson is on-shell in the decay $A^0\to h^0 Z/H^0 Z$ with
$m_{h^0/H^0}=125$ GeV.
However, the above cut on the invariant mass of the $b\overline b$ pair (which was chosen to be 100 GeV$ < m_{b\overline b} < 140$ GeV) would strongly suppress the number of events arising from $h^0$ with $m_{h^0}< 125$ GeV in IH, as emphasised in our
previous work in \cite{Akeroyd:2023kek}. However, we erroneously stated in \cite{Akeroyd:2023kek} that there had been no search at the LHC for $A^0\to h^0 Z$ that is sensitive to $m_{h^0}<125$ GeV, but the CMS search in  \cite{CMS:2019ogx} does 
set limits on $m_{h^0}<125$ GeV.
 Consequently, we now summarise in some detail the search strategy and results of such an analysis, which was not known to us at the time of writing our article Ref.~\cite{Akeroyd:2023kek}.
 We note that the related decay $H^0\to A^0 Z$ is also searched for in   \cite{CMS:2019ogx}, but we will only consider $A^0\to h^0Z^{(*)}$. 
 We emphasise that no ATLAS search for $A^0\to h^0 Z$ has probed the region $m_{h^0}<125$ GeV, with the ATLAS searches in  \cite{ATLAS:2020gxx,ATLAS:2022enb,ATLAS:2023zkt} all being for the NH scenario of  $m_{H^0}>125$ GeV.
 \begin{table}[h]
\begin{center}
\begin{tabular}{|c||c|c|}
\hline
& Invariant mass $m_{b\overline b}$ & Invariant mass $m_{l^+l^-}$  \\ \hline
CMS, 13 TeV, 35.9 fb$^{-1}$ \cite{CMS:2019qcx} 
& 100 GeV$ < m_{b\overline b} < 140$ GeV  &  70 GeV$ < m_{l^+l^-} < 110$ GeV\\
CMS, 13 TeV, 35.9 fb$^{-1}$  \cite{CMS:2019ogx} 
&  no cut & 70 GeV$ < m_{l^+l^-} < 110$ GeV \\
\hline
\end{tabular}
\end{center}
\caption{CMS Searches for $A^0\to h^0/H^0 Z$ at the LHC with 13 TeV data.}
\label{LHC_search}
\end{table}
The search in \cite{CMS:2019ogx}  differs from the search in \cite{CMS:2019qcx} by targeting the decays $A^0\to h^0 Z$ and $A^0\to H^0 Z$ over a wide range of masses of $h^0/H^0$, and includes the
case of $m_{h^0,H^0}=125$ GeV. In the notation of  Ref.~\cite{CMS:2019ogx} the discovered 125 GeV boson is always denoted by $h^0$, while $H^0$ can be lighter or heavier than $h^0$ and is taken as
a free parameter in the range 30 GeV $< m_{H^0} <1000$ GeV. The restriction that 
$m_{A^0} > m_{H^0} + m_Z$ (i.e. to ensure an on-shell $Z$ in the decay  $A^0\to H^0 Z$) means that results are shown for the range 30 GeV 
$< m_{H^0} < 900$ GeV with the heavier $A^0$ being in the range 120 GeV 
$< m_{A^0} < 1000$ GeV.  The search strategy in Ref.~\cite{CMS:2019ogx} does not employ a cut on the invariant mass of the $b\overline b$ pair (in difference to \cite{CMS:2019qcx}) as the aim is to be sensitive to a wide range of masses
of $m_{h^0}$ and $m_{H^0}$.
We note that in our work the IH scenario is defined 
by the (different) notation $m_{H^0}=125$ GeV and $m_{h^0} < 125$ GeV. We will use our notation in what follows.

In the search in \cite{CMS:2019ogx} the region of $m_{h^0}<125$ GeV corresponds to the IH scenario and is probing the decay 
$A^0\to h^0 Z$. The case of $m_{H^0}>125$ GeV is for the NH scenario and is probing the decay $A^0\to H^0 Z$. The region of $m_{h^0,H^0}=125$ GeV (NH and IH) is probed in \cite{CMS:2019ogx} without any cut on
 the invariant mass of the $b\overline b$ pair. Consequently, the limit on the cross section for masses of  $m_{h^0,H^0}=125$ GeV is weaker that that in  \cite{CMS:2019qcx} 
 (in the latter the cut was chosen to be 100 GeV$ < m_{b\overline b} < 140$ GeV, as mentioned earlier). We do not consider the case of (approximately) degenerate CP-even scalars $m_{h^0}\approx m_{H^0}=125$ GeV.
For the IH case on which we shall focus one has $m_{H^0}=125$ GeV and  $m_{h^0}<125$ GeV. Hence the search strategy in  \cite{CMS:2019ogx} could give signals
for both $A^0\to h^0 Z$ and $A^0\to H^0 Z$, although the former has a much larger rate due to the larger mass difference between $m_{A^0}$ and $m_{h^0}$ and the fact that 
in the IH scenario the coupling $A^0h^0Z\sim \cos(\beta-\alpha)\approx 1$ while $A^0H^0Z\sim \sin(\beta-\alpha)\approx 0$.

The signal in  \cite{CMS:2019ogx} is $A^0\to h^0Z/H^0Z\to b\overline b l^+l^-$, where $l^+l^-=e^+e^-$ or $\mu^+\mu^-$.
The main background is $Z$ boson production through the Drell Yan process, with the $t\overline t$ background being next in importance. Much smaller contributions are from single $t$, diboson $VV$ (where $V=Z,W^\pm$), triboson $VVV$, and others. 
Signal selection involves transverse momentum
cuts on  $l^+l^-$ and two tagged $b$ jets are required (with a $b$-tagging efficiency of 70\%).
As the $Z$ boson in $A^0\to h^0Z/H^0Z$ is taken to be on-shell, a cut on the invariant mass of the leptons, ${\rm 70 \,GeV} < m_{l^+l^-}<110$ GeV, is employed to reduce the backgrounds without $ZZ$ production
while keeping most of the signal. The invariant mass ($m_{b\overline b}$) of $b\overline b$ from the decay of $h^0/H^0$ will be peaked at $m_{h^0,H^0}$ and the invariant mass 
($m_{b\overline b l^+l^-}$) of $b\overline b l^+l^-$ from $A^0$  will be peaked at $m_{A^0}$. These two invariant masses are used to discriminate the signal from the background. Excluded
cross sections for $A^0\to h^0Z/H^0Z\to b\overline b l^+l^-$ are shown in the plane $[m_{A^0},m_{h^0,H^0}]$ ranging from $> 1000$ fb (the weakest limits) for light masses (e.g. $m_{A^0}=150$ GeV and $m_{h^0}=30$ GeV)
to $>1$ fb  (e.g. $m_{A^0}=900$ GeV and $m_{H^0}=400$ GeV). These limits are then interpreted in the 2HDM (Type II) to give excluded regions in the plane $[m_{A^0},m_{h^0,H^0}]$ for fixed values of $\cos(\beta-\alpha)$ and $\tan\beta$, and
excluded regions in the plane  $[\cos(\beta-\alpha),\tan\beta]$ for fixed values of $m_{A^0}$ and $m_{h^0,H^0}$. 
Importantly, there are no limits for $A^0\to h^0Z^*/H^0Z^*$ (i.e. for an off-shell $Z^*$ when $m_{A^0}-m_{h^0,H^0}< m_Z$) because the
aforementioned cut ${\rm 70 \,GeV} < m_{l^+l^-}<110$ GeV would remove most of these events.

\section{Results}
\noindent
In this section we show our results for the following signal cross sections in the IH scenario in the four types of 2HDM with NFC:
\begin{equation}
\sigma(gg\to A^0\to b\overline b Z^{(*)})=\sigma(gg\to A^0)\times {\rm BR}(A^0\to h^0Z^{(*)})\times {\rm BR}(h^0\to b\overline b)\,,
\label{event_number_bb}
\end{equation}
and
\begin{equation}
\sigma(gg\to A^0\to \tau^+\tau^- Z^{(*)})=\sigma(gg\to A^0)\times {\rm BR}(A^0\to h^0Z^{(*)})\times {\rm BR}(h^0\to \tau^+ \tau^-)\,.
\label{event_number_tautau}
\end{equation}
Note that we have not included the additional BR suppression arising from the decays of $Z^{(*)}$ (i.e. $Z^{(*)}\to e^+e^-,\mu^+\mu^-$, each with a BR of $\approx 0.03$) which is included in 
the presented limits on $\sigma(gg\to A^0\to b\overline b Z)$ in \cite{CMS:2019ogx}. In order to compare $\sigma(gg\to A^0\to b\overline b Z^{(*)})$ with the limits
in \cite{CMS:2019ogx} it is necessary to multiply by the sum of these BRs ($\approx 0.06$).

In our previous work \cite{Akeroyd:2023kek} the magnitude of $\sigma(gg\to A^0\to b\overline b Z^{(*)})$ alone was studied in the 2HDM (Type I) and we were not aware of
the search in \cite{CMS:2019ogx} in which limits on the cross section are obtained for $m_{h^0}<125$ GeV. In this work we
extend the study of \cite{Akeroyd:2023kek} in the following ways:

\begin{itemize}
\item[{(i)}]  The magnitude of $\sigma(gg\to A^0\to b\overline b Z^{(*)})$  and $\sigma(gg\to A^0\to \tau^+\tau^- Z^{(*)})$ is studied in 
all four 2HDMs with NFC in the IH scenario.

\item[{(ii)}]  The constraints of the search \cite{CMS:2019ogx} are respected by making use of HiggsTools \cite{Bahl:2022igd}.

\item[{(iii)}] We emphasise that the region of $m_{A^0} < 200$ GeV and $m_{h^0}<125$ GeV (of interest to us) that is probed in the search in  \cite{CMS:2019ogx} has weaker limits on the cross section than for
the case of $m_{A^0} > 200$ GeV. Importantly, the search strategy in \cite{CMS:2019ogx}  focusses only on the case of the $Z$ boson being on-shell in the decay $A^0\to h^0 Z$, and 
does not present limits for the case of the $Z$ boson being off-shell (i.e. $A^0\to h^0 Z^*$). This leads to a sizeable unexcluded region in the plane $[m_{h^0,H^0},m_{A^0}]$. We suggest cuts 
on the invariant masses of the leptons ($m_{l\overline l}$) and $b$ quarks ($m_{b\overline b}$) that might improve sensitivity to $\sigma(gg\to A^0\to b\overline b Z^{*})$
and $\sigma(gg\to A^0\to \tau^+\tau^- Z^{*})$.

\end{itemize}

We note again that at the time of writing of \cite{Akeroyd:2023kek} we were not aware of \cite{Bernon:2015wef} in which the magnitude of $\sigma(gg\to A^0\to b\overline b Z^{(*)})$ in the IH scenario was studied in the 2HDM (Type I and Type II), but not in
the flipped and lepton specific types. However, the presentation of results is different in \cite{Bernon:2015wef}  and \cite{Akeroyd:2023kek}. In \cite{Bernon:2015wef}  the magnitude of  $\sigma(gg\to A^0\to h^0 Z^{(*)})$ without the factor BR$(h^0\to b\overline b)$ 
was studied as a function of $m_{A^0}$ via a scatter plot in which $m_{h^0}$ is varied in the range $10\,{\rm GeV} < m_{h^0} < 121.5\, {\rm GeV}$ and other 2HDM parameters were varied in ranges.
 Consequently, the dependence on $m_{h^0}$ is somewhat hidden. 
In contrast, in  \cite{Akeroyd:2023kek} (and in the present work) all 2HDM parameters except $m_{A^0}$ are fixed,  and $\sigma(gg\to A^0\to b\overline b Z^{(*)})$ is displayed 
for three values of $m_{h^0}$. We will reproduce and update the results for $\sigma(gg\to A^0\to h^0 Z^{(*)})$ in \cite{Bernon:2015wef}  by performing a scan over the same region of parameter space but with the
latest version of HiggsTools.

Having clarified how the present work goes beyond what was done in \cite{Bernon:2015wef} and  \cite{Akeroyd:2023kek}, we now discuss the dependence of $\sigma(gg\to A^0\to b\overline b Z^{(*)})$ in 
eq.~(\ref{event_number_bb}) and $\sigma(gg\to A^0\to \tau^+\tau^- Z^{(*)})$ in eq.~(\ref{event_number_tautau}) on the 2HDM parameters (which was also discussed in  \cite{Akeroyd:2023kek}).
In NH these cross sections depend on three unknown parameters: $m_{A^0}$, $\tan\beta$ and $\cos(\beta-\alpha)$.
 In IH there is a fourth unknown parameter, $m_{h^0}$. The dependence of the three terms in eq.~(\ref{event_number_bb}) and eq.~(\ref{event_number_tautau})  on the 2HDM parameters is as follows:
 \begin{itemize}
\item[{(i)}]  the cross-section $\sigma(gg\to A^0)$ depends on $m_{A^0}$ and $\tan\beta$ via the Yukawa couplings $y^t_{A^0}$  and $y^b_{A^0}$ (see Table I, for the $\tan\beta$ dependence in each 2HDM). 
Contributions from the 
couplings of $A^0$ to lighter fermions can be neglected due to their much smaller masses;
\item[{(ii)}]  BR$(A^0\to h^0Z^{(*)})$ is given by $\Gamma(A^0\to h^0Z^{(*)})/\Gamma^{\rm total}_{A^0}$. 
The partial width $\Gamma(A^0\to h^0Z^{(*)})$ depends on $m_{A^0}$, the mass difference $m_{A^0}-m_{h^0}$ (in the phase space factor) and $\cos^2(\beta-\alpha)$ (in the square of the $A^0h^0Z$ coupling).
The total width $\Gamma^{\rm total}_{A^0}$ is equal to  $\Gamma(A^0\to h^0Z^{(*)})+\Gamma^{\rm rest}_{A^0}$, where $\Gamma^{\rm rest}_{A^0}$ is the sum of the partial decay widths of 
all the other decays of $A^0$;
\item[{(iii)}]   BR$(h^0\to b\overline b$) is given by $\Gamma(h^0\to b\overline b)/\Gamma^{\rm total}_{h^0}$.
The partial width $\Gamma(h^0\to b\overline b)$ depends on $m_{h^0}$ and $\cos^2(\beta-\alpha)$ (via the scaling factors  $\sin\alpha/\cos\beta$ and $\cos\alpha/\sin\beta$). 
The total width $\Gamma^{\rm total}_{h^0}$ is equal to  $\Gamma(h^0\to b\overline b)+\Gamma^{\rm rest}_{h^0}$, where $\Gamma^{\rm rest}_{h^0}$ is the sum of the partial decay widths of 
all the other decays of $h^0$.  Analogous comments apply to  BR$(h^0\to \tau^+\tau^-$).
\end{itemize}

\begin{figure}
    \centering
    \includegraphics [width=17cm]{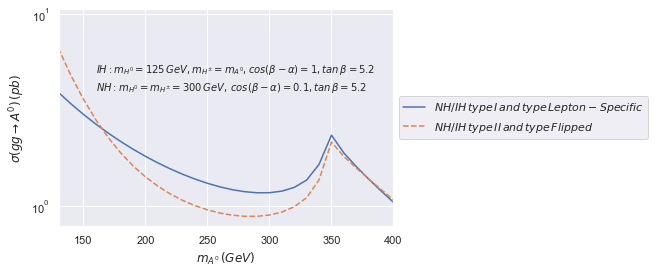}
    \caption{The cross section $\sigma(gg\to A^0)$ as a function of $m_{A^0}$ for the 2HDM (Type I and Lepton Specific) in NH/IH, and 
   for the 2HDM (Type II and Flipped) in NH/IH. The values of the input parameters are
    displayed on the figure.}
    \label{fig:sig_mA} 
    \end{figure}

We are now present the novel results of this work.
In Fig.~\ref{fig:sig_mA} the cross section $\sigma(gg\to A^0)$ is plotted as a function of $m_{A^0}$ for the 2HDM (Type I and Lepton Specific) in NH/IH, and 
   for the 2HDM (Type II and Flipped) in NH/IH.  In \cite{Akeroyd:2023kek} the analogous plot did not show $\sigma(gg\to A^0)$ in the Flipped and Lepton specific models.
   The cross section depends solely on $m_{A^0}$ and $\tan\beta$, and is the same in IH/NH for a particular 2HDM for specific values of these two parameters. We choose $\tan\beta=5.2$ in both NH and IH,
   which gives a different value of $\alpha$ for our input values of $\cos(\beta-\alpha)=0.1$ for NH and $\cos(\beta-\alpha)=1$ for IH.
In the 2HDM (Type I and Lepton Specific), $\sigma(gg\to A^0)$ is the same as these two models have identical $y^t_{A^0}$ and $y^b_{A^0}$. 
In the 2HDM (Type II and Flipped), $\sigma(gg\to A^0)$ is also the same as these two models have identical $y^t_{A^0}$ and $y^b_{A^0}$, but the dependence on $m_{A^0}$ is different to
that in the 2HDM (Type I and Lepton Specific) due to the different Yukawa couplings. 

\begin{figure}
    \centering
    \includegraphics [width=17cm]{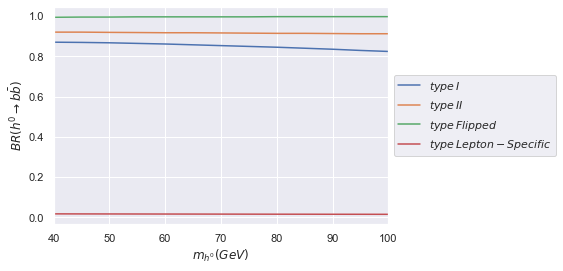}
    \caption{BR$(h^0\to b\overline b$) in the four types of 2HDM in IH as a function of $m_{h^0}$. The values of the  2HDM input parameters are given in the text.}
    \label{fig:BRbb}
\end{figure}

In Fig.~\ref{fig:BRbb}, BR$(h^0\to b\overline b$) is plotted as a function of $m_{h^0}$ in the four types of 2HDM in the IH scenario. In \cite{Akeroyd:2023kek} the analogous plot only showed
BR$(h^0\to b\overline b$) in the 2HDM (Type I). The 2HDM parameters are taken to be 
$m_{H^\pm}=m_{A^0}=400$ GeV, $m_{H^0}=125$ GeV, $\cos(\beta-\alpha)=1$, $\tan\beta=5.2$ and $m^2_{12}=m^2_{h^0}\tan\beta[1/(1+\tan^2\beta)]$. One can see that BR$(h^0\to b\overline b$) is between $80\%$
 and $\approx 100\%$ in the 2HDMs Type I, Type II and Flipped, which is considerably larger than BR$(H^0\to b\overline b)\approx 58\%$ for the SM Higgs boson. This is because
 the competing decay $h^0\to W^{(*)}W^*$ is significantly suppressed by phase space in the plotted range of 40 GeV $< m_{h^0}<100$ GeV compared to a Higgs boson of 125 GeV.
 The relative importance of BR$(h^0\to b\overline b$) being Flipped $>$ Type II $>$ Type I is due to the different forms of the Yukawa couplings in these 2HDMs.
 In contrast, for the 2HDM (Lepton Specific), BR$(h^0\to b\overline b$) is only a few percent and this is due to the fact that 
 the coupling $h^0\tau^+\tau^-$ is enhanced relative to the $h b\overline b$ coupling in this model.

\begin{figure}
    \centering
    \includegraphics [width=17cm]{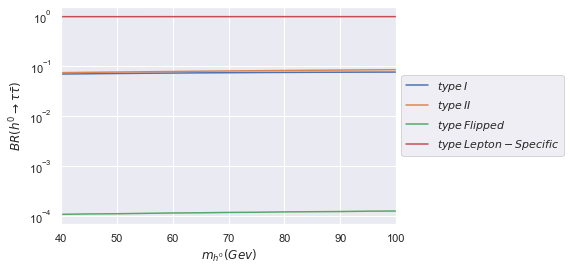}
    \caption{BR$(h^0\to \tau^+\tau^-$) in the four types of 2HDM in IH as a function of $m_{h^0}$. The values of the  2HDM input parameters are given in the text.}
    \label{fig:BRtautau}
\end{figure}

In Fig.~\ref{fig:BRtautau}, BR$(h^0\to \tau^+\tau^-$) in the four types of 2HDM in the IH scenario is plotted as a function of $m_{h^0}$. In \cite{Akeroyd:2023kek} the analogous plot only showed
BR$(h^0\to \tau^+\tau^-$) in the 2HDM (Type I).  The 2HDM parameters are fixed as in 
Fig.~\ref{fig:BRbb}. As expected, one sees an opposite result to that in Fig.~\ref{fig:BRbb}, with BR$(h^0\to \tau^+\tau^-$) close to $100\%$ in the Lepton Specific model, being around $10\%$ in
Type I and Type II, and less than a percent in the Flipped model.

\begin{figure}
    \centering
    \includegraphics [width=17cm]{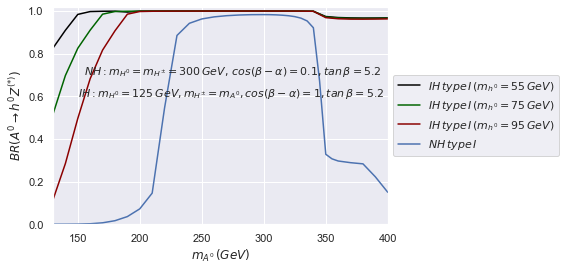}
    \caption{BR$(A^0\to h^0Z^{(*)})$ in NH and IH as a function of $m_{A^0}$ in the 2HDM (Type I). The values of the 2HDM input parameters are
    displayed on the figure. }
    \label{fig:BRAhZ_type1}
\end{figure}

\begin{figure}
    \centering
    \includegraphics [width=17cm]{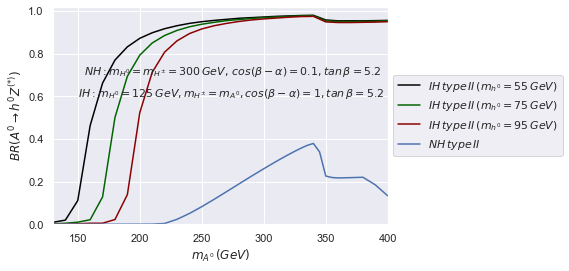}
    \caption{BR$(A^0\to h^0Z^{(*)})$ in NH and IH as a function of $m_{A^0}$ in the 2HDM (Type II). The values of the 2HDM input parameters are
    displayed on the figure.}
    \label{fig:BRAhZ_type2}
\end{figure}

\begin{figure}
    \centering
    \includegraphics[width=17cm]{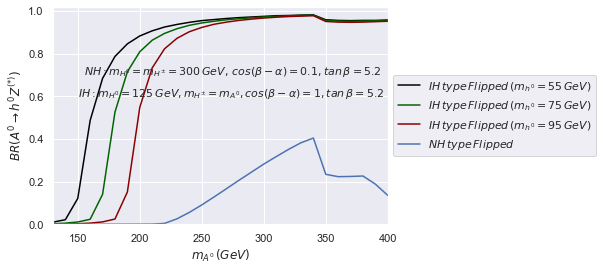}
    \caption{BR$(A^0\to h^0Z^{(*)})$ in NH and IH as a function of $m_{A^0}$ in the 2HDM (Flipped). The values of the input parameters are
    displayed on the figure.}
    \label{fig:BRAhZ_typeF}
    \end{figure}

\begin{figure}

    \centering
    \includegraphics[width=17cm]{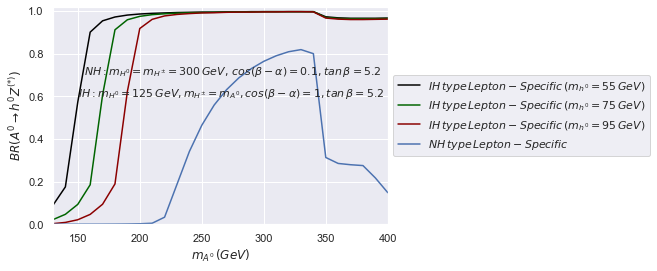}
    \caption{BR$(A^0\to h^0Z^{(*)})$ in NH and IH as a function of $m_{A^0}$ in the 2HDM (Lepton Specific). The values of the input parameters are
    displayed on the figure.}
    \label{fig:BRAhZ_typeLS}
\end{figure}

In Fig.~\ref{fig:BRAhZ_type1}, Fig.~\ref{fig:BRAhZ_type2}, Fig.~\ref{fig:BRAhZ_typeF} and Fig.~\ref{fig:BRAhZ_typeLS}, BR$(A^0\to h^0Z^{(*)})$ in NH and IH is plotted as a function of $m_{A^0}$ in the 2HDM (Type I),
2HDM (Type II), 2HDM (Flipped) and 2HDM (Lepton Specific) respectively. The values of the input parameters are
    displayed on the figure, and in the IH scenario are the same as in Fig.~\ref{fig:BRbb} and Fig.~\ref{fig:BRtautau} except for $m_{A^0}$ (which is varied from 130 GeV to 400 GeV, while taking
    $m_{H^\pm}=m_{A^0}$) and  $m_{h^0}$, which is fixed
    to 55 GeV, 75 GeV and 95 GeV in IH.  In \cite{Akeroyd:2023kek} the analogous plot only showed  BR$(A^0\to h^0Z^{(*)})$
 in the 2HDM (Type I). 
 In IH, for a given value of $m_{A^0}$ and $m_{h^0}$ one can see that BR$(A^0\to h^0Z^{(*)})$ has its largest value in Type I, followed by Lepton Specific, with Type II and Flipped having the smallest
 (and roughly the same) values e.g. for $m_{A^0}=150$ GeV and $m_{h^0}=55$ GeV, one has BR$(A^0\to h^0Z^{(*)})\approx 90\%,10\%,10\%$ and $60\%$ in Type I, Type II, Flipped
and Lepton Specific, respectively. In NH, BR$(A^0\to h^0Z^{(*)})$ is usually much smaller for a given value of $m_{A^0}$ due to the (necessary) smaller value of $\cos(\beta-\alpha)$ and the smaller mass splitting $m_{A^0}-m_{h^0}$ for $m_{h^0}=125$ GeV.
Note that in Type I and NH, BR$(A^0\to h^0Z^{(*)})$ can be close to $100\%$ when the $Z$ boson is on-shell. Importantly, as emphasised in \cite{Akeroyd:2023kek} (and originally pointed out in \cite{Akeroyd:1998dt}),
BR$(A^0\to h^0Z^*)$ can be very large even with an off-shell $Z^*$.

\begin{figure}
    \centering
    \includegraphics[width=17cm]{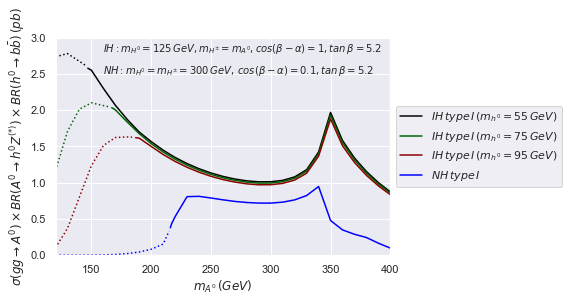}
    \caption{$\sigma(gg\to A^0)\times {\rm BR}(A^0\to h^0Z^{(*)})\times {\rm BR}(h^0\to b\overline b)$ in NH and IH as a function of $m_{A^0}$ in the 2HDM (Type I). The values of the input parameters are
    displayed on the figure. The solid (dotted) line denotes on-shell $Z$ (off-shell $Z^*$).}
    \label{fig:SigXBRXBRbb_type1}
\end{figure}

\begin{figure}
    \centering
    \includegraphics[width=17cm]{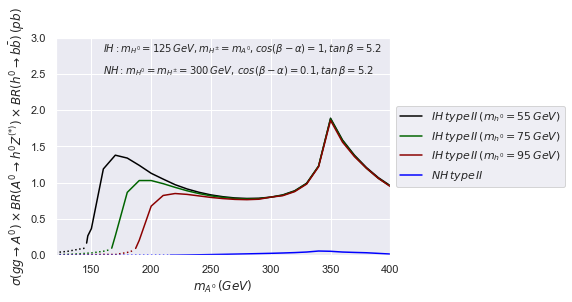}
    \caption{$\sigma(gg\to A^0)\times {\rm BR}(A^0\to h^0Z^{(*)})\times {\rm BR}(h^0\to b\overline b)$ in NH and IH as a function of $m_{A^0}$ in the 2HDM (Type II). The values of the input parameters are
    displayed on the figure. The solid (dotted) line denotes on-shell $Z$ (off-shell $Z^*$).}
    \label{fig:SigXBRXBRbb_type2}
\end{figure}

\begin{figure}
    \centering
    \includegraphics[width=17cm]{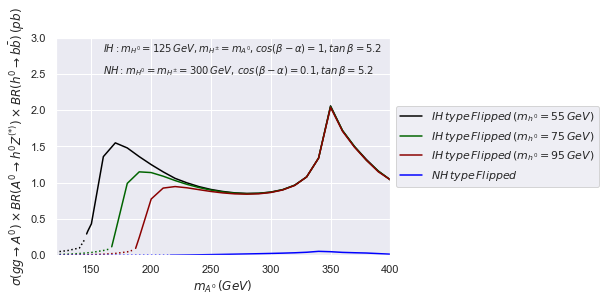}
    \caption{$\sigma(gg\to A^0)\times {\rm BR}(A^0\to h^0Z^{(*)})\times {\rm BR}(h^0\to b\overline b)$ in NH and IH as a function of $m_{A^0}$ in the 2HDM (Flipped). The values of the input parameters are
    displayed on the figure. The solid (dotted) line denotes on-shell $Z$ (off-shell $Z^*$).}
    \label{fig:SigXBRXBRbb_typeF}
\end{figure}

\begin{figure}
    \centering
    \includegraphics[width=17cm]{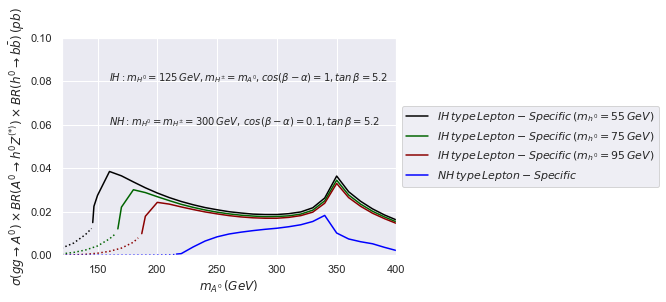}
    \caption{$\sigma(gg\to A^0)\times {\rm BR}(A^0\to h^0Z^{(*)})\times {\rm BR}(h^0\to b\overline b)$ in NH and IH as a function of $m_{A^0}$ in the 2HDM (Lepton Specific). The values of the input parameters are
    displayed on the figure. The solid (dotted) line denotes on-shell $Z$ (off-shell $Z^*$).}
    \label{fig:SigXBRXBRbb_typeLS}
\end{figure}

  In Fig.~\ref{fig:SigXBRXBRbb_type1}, Fig.~\ref{fig:SigXBRXBRbb_type2},  Fig.~\ref{fig:SigXBRXBRbb_typeF} and Fig.~\ref{fig:SigXBRXBRbb_typeLS}
 the product that determines the event number in the LHC search in  \cite{CMS:2019ogx}, being $\sigma(gg\to A^0)\times {\rm BR}(A^0\to h^0Z^{(*)})\times {\rm BR}(h^0\to b\overline b)$ in 
 eq.~(\ref{event_number_bb}), is plotted in NH and IH as a function of $m_{A^0}$ in 
 in the 2HDM (Type I), 2HDM (Type II), 2HDM (Flipped) and 2HDM (Lepton Specific) respectively. The input parameters are the same as in Fig.~\ref{fig:BRAhZ_type1}, Fig.~\ref{fig:BRAhZ_type2}, Fig.~\ref{fig:BRAhZ_typeF} and Fig.~\ref{fig:BRAhZ_typeLS}.
 The $m_{A^0}$ dependence can be understood from the previous plots e.g. the local peak at $m_{A^0}\approx 350\,{\rm GeV}=2m_t$ is from the increase in $\sigma(gg\to A^0)$ while
 the local peak at lower values of $m_{A^0}$ is when the rise in BR$(A^0\to h^0Z^{(*)})$ is offset by the fall in $\sigma(gg\to A^0)$. On all figures the dotted line indicates the values of $m_{A^0}$ 
 for which the $Z$ is off-shell, while the solid line corresponds to the case of an on-shell $Z$. 
 In the 2HDM (Type I, Type II and Flipped) the signal cross section can reach a few pb, while in the 2HDM (Lepton Specific) the suppression factor from BR$(h^0\to \tau^+\tau^-)$ is severe and the signal cross section reaches a maximum of
 0.04 pb. We do not include the suppression factor of BR$(Z^*\to l^+l^-)\approx 0.06$ which is required when comparing to the excluded cross sections for the $b\overline b l^+l^-$ signal
 in \cite{CMS:2019ogx}.
 
\begin{figure}
    \centering
    \includegraphics[width=17cm]{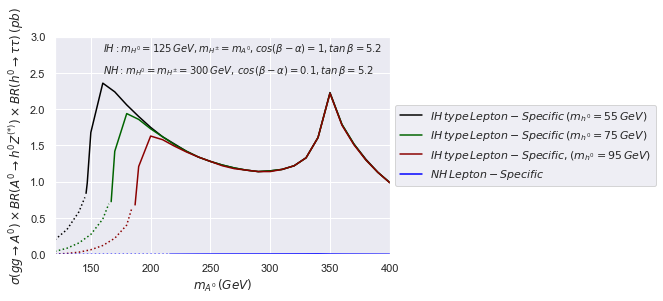}
    \caption{$\sigma(gg\to A^0)\times {\rm BR}(A^0\to h^0Z^{(*)})\times {\rm BR}(h^0\to \tau^+\tau^-)$ in NH and IH as a function of $m_{A^0}$ in the 2HDM (Lepton Specific). The values of the input parameters are
    displayed on the figure. The solid (dotted) line denotes on-shell $Z$ (off-shell $Z^*$).}
    \label{fig:SigXBRXBRtautau_typeLS}
\end{figure}

In Fig.~\ref{fig:SigXBRXBRtautau_typeLS} the product $\sigma(gg\to A^0)\times {\rm BR}(A^0\to h^0Z^{(*)})\times {\rm BR}(h^0\to \tau^+\tau^-)$ is plotted in NH and IH as a function of $m_{A^0}$ in the
 2HDM (Lepton Specific), for which
BR$(h^0\to \tau^+\tau^-)$ is large. We do not show results for the other three models as BR$(h^0\to \tau^+\tau^-)$ is small. The signal cross section can be as large as 2 pb and would be the best way to
probe the 2HDM (Lepton Specific) with $A^0\to h^0Z^{(*)}$ decay.
There has been a search at the LHC with 8 TeV data (but not 13 TeV) for the $h^0\to \tau^+\tau^-$ in this channel \cite{CMS:2016xnc}. 

Finally, we study the magnitude of $\sigma(gg\to A^0)\times {\rm BR}(A^0\to h^0Z^{(*)})$ without the factor BR($h\to b\overline b)$ in IH with a scan of the 2HDM parameter space in the following ranges:
\begin{eqnarray}
-\pi/2 \le \alpha \le \pi/2,\;\;   0.5 \le \tan\beta \le 60,\;\;   -(2000 \,{\rm GeV})^2 \le m^2_{12} \le (2000\, {\rm GeV})^2,\;\;  \nonumber \\
10 \,{\rm GeV} \le m_{h^0} \le 121.5 \,{\rm GeV},\;\; 5 \,{\rm GeV} \le m_{A^0} \le 2000 \,{\rm GeV},\;\;m^* \le m_{H^\pm} \le 2000 \,{\rm GeV},
\label{2HDMscan}
\end{eqnarray}
where $m^*$ is the lower bound on $m_{H^\pm}$ in Type I or Type II, taken to be 300 GeV and 600 GeV respectively. This is the same parameter space for the scan in \cite{Bernon:2015wef} in which 
the magnitude of $\sigma(gg\to A^0)\times {\rm BR}(A^0\to h^0Z^{(*)})$ was also studied. As mentioned earlier in this section, at the time of writing of \cite{Akeroyd:2023kek} we were 
unaware of the work of \cite{Bernon:2015wef}, but the presentation of results is different in   \cite{Akeroyd:2023kek}  (showing $m_{h^0}$ dependence explicitly) and  \cite{Bernon:2015wef} 
(not showing $m_{h^0}$ dependence explicitly). We now update the results of \cite{Bernon:2015wef} using HiggsTools \cite{Bahl:2022igd} to constrain the 2HDM parameter space in eq.~(\ref{2HDMscan}).
We also note that the LHC search for $A^0\to h^0Z$ with 8 TeV  \cite{CMS:2016xnc} was used to constrain  $\sigma(gg\to A^0)\times {\rm BR}(A^0\to h^0Z^{(*)})$ in  \cite{Bernon:2015wef}.

\begin{figure}
    \centering
    \includegraphics[width=17cm]{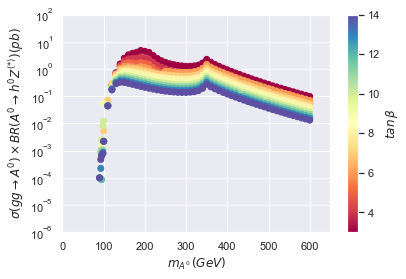}
    \caption{$\sigma(gg\to A^0)\times {\rm BR}(A^0\to h^0Z^{(*)})$ as a function of $m_{A^0}$ with all other 2HDM parameters varied in the ranges given in the text. The dependence on $\tan\beta$ is shown explicitly
    as a colour code.}
    \label{fig:SigXBRscan_tanb_type1}
\end{figure}

\begin{figure}
    \centering
    \includegraphics[width=17cm]{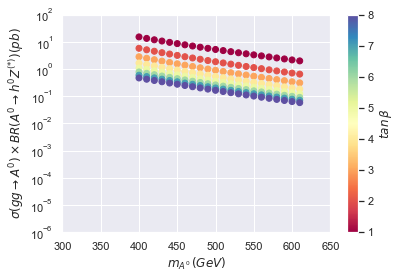}
    \caption{ $\sigma(gg\to A^0)\times {\rm BR}(A^0\to h^0Z^{(*)})$ as a function of $m_{A^0}$ with all other 2HDM parameters varied in the ranges given in the text. The dependence on $\tan\beta$ is shown explicitly as a colour code.}
    \label{fig:SigXBRscan_tanb_type2}
\end{figure}

In Fig.~\ref{fig:SigXBRscan_tanb_type1}, $\sigma(gg\to A^0)\times {\rm BR}(A^0\to h^0Z^{(*)})$ is plotted as a function of $m_{A^0}$ in the 2HDM (Type I) with all other 2HDM parameters varied in the ranges given in  eq.~(\ref{2HDMscan}).
The $\tan\beta$ dependence is given by the colour coding, with the largest cross sections arising for low $\tan\beta$ (as expected in Type I). We find good agreement with Fig.~19 in \cite{Bernon:2015wef} e.g. the maximum cross sections are of the order of a few pb, with the peak occurring at around $m_{A^0}=180$ GeV, and no points being
found with $m_{A^0}<80$ GeV.
However, a notable difference is the allowed range of $\tan\beta$. In  \cite{Bernon:2015wef} points are found in the range $0.5 \le \tan\beta \le 60$, with the points at very low $\tan\beta$ ($<1)$ being restricted to $m_{A^0}>300$ GeV. In contrast, in
Fig.~\ref{fig:SigXBRscan_tanb_type1} the range of values of $\tan\beta$ is $3 \le \tan\beta \le 14$. In \cite{Bernon:2015wef} for $m_{A^0}=200$ GeV, points with $\tan\beta$ just below 2 are found, while in Fig.~\ref{fig:SigXBRscan_tanb_type1} 
for $m_{A^0}=200$ GeV the lowest value of $\tan\beta$ is 3.
 Fig.~\ref{fig:SigXBRscan_tanb_type2} is the same as Fig.~\ref{fig:SigXBRscan_tanb_type1} but for the 2HDM (Type II). Again, we find good agreement with the corresponding plot in \cite{Bernon:2015wef}, with
 differences being the allowed range of $\tan\beta$, which is $1 \le \tan\beta \le 8$ in Fig.~\ref{fig:SigXBRscan_tanb_type2} and $0.5 \le \tan\beta \le 50$ in \cite{Bernon:2015wef}, and the lowest value of $m_{A^0}$ (being
 $m_{A^0}=400$ GeV in Fig.~\ref{fig:SigXBRscan_tanb_type2}  and $m_{A^0}=430$ GeV in  \cite{Bernon:2015wef}).
 
Having shown all our numerical results we now make an important remark on the LHC search for  $\sigma(gg\to A^0)\times {\rm BR}(A^0\to h^0Z^{(*)})\times {\rm BR}(h^0\to b\overline b)$ in \cite{CMS:2019ogx}, in which
a cut on the invariant mass of the leptons $70\,{\rm GeV}< m_{l^+l^-} < 110$ GeV is imposed.
This cut ensures that backgrounds in which $l^+l^-$ do not originate from $Z$ bosons (e.g. $t\overline t$) are suppressed, while the signal $A^0\to h^0 Z$
is largely unaffected by this cut provided that $Z$ is on-shell in this decay. However, this cut on $m_{l^+l^-}$ would remove a large fraction (and possibly all)  of signal events $A^0\to h^0Z^*$ and hence
  no limits are presented in  \cite{CMS:2019ogx}  for the case of the $Z$ boson being off-shell.
This leads to a sizeable unexcluded region in the plane $[m_{h^0,H^0},m_{A^0}]$ i.e. the region where $m_{A^0}-m_{h^0} < m_Z$ (with $m_{A^0}>m_{h^0}$). Importantly, as shown in Fig.~\ref{fig:SigXBRXBRbb_type1}, Fig.~\ref{fig:SigXBRXBRbb_type2},  Fig.~\ref{fig:SigXBRXBRbb_typeF} and Fig.~\ref{fig:SigXBRXBRbb_typeLS},
the signal cross section can still be sizeable for the case of $A^0\to h^0 Z^*$, especially in Type I. In order to provide sensitivity to the case of $Z^*$ we suggest
relaxing the range of the invariant mass cut on the leptons e.g. using $10\,{\rm GeV}< m_{l^+l^-} < 110$ GeV instead of $70\,{\rm GeV}< m_{l^+l^-} < 110$ GeV. Such a cut would
capture more of the signal from $A^0\to h^0 Z^*$, but would also increase backgrounds (such as $t\overline t$). However, in order to improve sensitivity to the IH scenario 
with $A^0\to h^0 Z^*$, a cut on
the invariant mass ($m_{b\overline b}$) of $b\overline b$ could also be imposed (no such cut is used in  \cite{CMS:2019ogx}, as the aim there was to be sensitive to a wide region of values of $m_{h^0,H^0}$).
For example, the cut  $60\,{\rm GeV}< m_{b\overline b} < 100$ GeV would reduce all backgrounds while keeping most of the signal for values of $m_{h^0}$ in the middle of that range.
We suggest that the search strategy in  \cite{CMS:2019ogx} be generalised to include these (or similar) cuts on $m_{l^+l^-}$ and $m_{b\overline b}$ in order to quantify
the sensitivity to $A^0\to h^0 Z^*$.


\section{Conclusions}  
\noindent
Searches are being carried out at the LHC for the decay of the CP-odd scalar $A^0$ in 2HDMs with NFC in the channel
$A^0\to h^0 Z$, where $h^0$ is either the discovered 125 GeV Higgs boson (called the NH scenario) or is an undiscovered CP-even scalar with a mass of below 125 GeV (called the IH scenario).
In the IH scenario this decay channel would provide the opportunity of simultaneous discovery of two scalars. 
By extending and developing the work in \cite{Bernon:2015wef} and our previous work in \cite{Akeroyd:2023kek} we studied the
magnitude of $\sigma(gg\to A^0)\times {\rm BR}(A^0\to h^0Z^{(*)})\times {\rm BR}(h^0\to b\overline b,\tau\overline \tau)$ in the four types of 2HDMs with NFC.
The CMS searches for $A^0\to h^0 Z$ in \cite{CMS:2019ogx} (with $\sqrt s=13$ TeV data) and \cite{CMS:2016xnc} (with $\sqrt s=8$ TeV data) are at present the only searches to probe
the region $m_{h^0}<125$ GeV in this channel, and limits are set on the relevant 2HDM parameters  $[m_{A^0}, m_{h^0},\tan\beta, \cos(\beta-\alpha)]$.
However, in both searches the selection cuts are optimised for the case of an on-shell $Z$ boson. For the case of the $Z$ boson being off-shell
(denoted by $Z^*$), no limits are set on the above 2HDM parameters from this process.
It is known that the decay $A^0\to h^0 Z^*$ can have a large BR in 2HDMs (especially in the Type I structure).
We suggested selection cuts on the invariant masses of $l^+l^-$ (from $Z^{(*)}$) and $b\overline b$ (from $h^0$) that might provide sensitivity to the decay $A^0\to h^0Z^*$ in the IH scenario and
thus extend the coverage of the $[m_{A^0},m_{h^0}]$ plane to include the regions where $m_{A^0} - m_{h^0}<m_Z$. We encourage a simulation
of the detection prospects for this scenario.

\section*{Acknowledgements}
\noindent
SA acknowledges the use of the IRIDIS High Performance Computing Facility, and associated support services at the University of Southampton. SA also acknowledges
support from a scholarship of the Imam Mohammad Ibn Saud Islamic University.
AA and SM are funded in part through the STFC CG ST/L000296/1.
SM is funded in part through the NExT Institute.

\end{document}